\documentclass[10pt, conference, compsocconf]{IEEEtran}
\usepackage{rotating}

\newcommand{\ocite}[1]{~\cite{#1}}
\newcommand{\commentedout}[1]{}

\newcommand {\apgt} {\ {\raise-.5ex\hbox{$\buildrel>\over\sim$}}\ }
\newcommand {\aplt} {\ {\raise-.5ex\hbox{$\buildrel<\over\sim$}}\ }

\begin{document}
%
\title{Survey of Multiscale and Multiphysics Applications and Communities}


\author{\IEEEauthorblockN{Derek Groen, Stefan J. Zasada, Peter V. Coveney}
\IEEEauthorblockA{Centre for Computational Science\\
University College London\\
London, United Kingdom\\
E-mail: d.groen@ucl.ac.uk, p.v.coveney@ucl.ac.uk}
}
\maketitle

\begin{abstract}

Multiscale and multiphysics applications are now commonplace, and
many researchers focus on combining existing models to construct combined
multiscale models. Here we present a concise review of multiscale applications
and their source communities. We investigate the prevalence of multiscale
projects in the EU and the US, review a range of coupling toolkits
they use to construct multiscale models and identify areas where
collaboration between disciplines could be particularly beneficial.  We
conclude that multiscale computing has become increasingly popular in recent
years, that different communities adopt very different approaches to
constructing multiscale simulations, and that simulations on a length scale of
a few metres and a time scale of a few hours can be found in many of the
multiscale research domains. Communities may receive additional benefit from 
sharing methods that are geared towards these scales.
\end{abstract}


%
\IEEEpeerreviewmaketitle

\section{Introduction}

Many physical problems we seek to understand nowadays are complex in nature,
and consist of separate physical processes that each contribute to the problem
as a whole. These processes each take
place on a specific space scale or time scale. In biology for example, the interactions
between molecules typically take place on a space scale of several nanometers
and a time scale of a number of nanoseconds. However, the interactions
on the cellular level will require considerably larger space and time scales.
Many problems are historically investigated by modeling or simulating a
physical process in isolation, and from the outcome of that exercise, determining
its contribution to the overall (complex) physical problem. In the last two
decades a new approach has become widespread, where researchers
construct models and simulations that capture multiple physical processes. Each
of these processes operates on a different space or time scale, has the potential 
to influence other processes, and is represented by a {\em submodel}.  This approach
is now known as {\em multiscale modelling} or {\em multiscale simulation}.
Here we use the term multiscale modelling to refer to both the
multiscale modelling and simulation of physical problems, and the term {\em
multiscale application} to refer to the program used to do the modelling. In
turn, we use the term {\em subcode} to refer to the implementations of each 
submodel. 

\subsection{Multiphysics modelling}

When a model captures multiple physical processes, and each of these processes
capture a different type of physics, it is commonly referred to as {\em
multiphysics modelling} or {\em multiphysics simulation}. For example, a model
of a star cluster that resolves Newtonian gravitational interactions using one
submodel and the aging of stars using another is considered to be a
multiphysics submodel, even if these models were (hypothetically) to operate on
the same space and time scale. However, a star cluster model that uses two
different submodels for the Newtonian gravitational interaction of stars is
generally not considered to be multiphysics, even when these models may be
applied on a different space or time scale.

Multiscale and multiphysics modelling are therefore two different concepts, but 
they do have one prime commonality in that they both consist of a number 
of submodels which have been combined (or {\em coupled}). A major challenge in
multiscale as well as multiphysics modelling lies in coupling these submodels
such that the overall model is both accurate enough to be scientifically 
relevant and reproducible, and efficient enough to be executed conveniently 
by modern compute resources.


\subsection{Multiscale and multiphysics applications}

Multiscale and multiphysics applications are present in a wide range of
scientific and engineering communities. By its nature, multiscale modeling is
highly interdisciplinary, with developments occurring independently across
research domains. 
Here we review a range of multiscale applications and
communities that reside within different scientific domains. We describe
several major projects for each domain and present the results of our
investigation on the popularity of multiscale simulation and modeling. We find
that multiscale methods are adopted in hundreds of projects both in the EU and
US, and that the popularity of multiscale simulation and modeling has increased
considerably in recent years. 

We also illustrate approaches to construct multiscale simulations
in different scientific domains, and compare some of the characteristics of
the multiscale communities in these domains. Additionally, we present a
comparison between coupling toolkits, and point out potential areas where
interdisciplinary collaborations could be particularly beneficial. Within this
survey we cover many major multiscale simulation and modeling activities, 
but this review is by no means exhaustive. For readability reasons we provide
only a limited number of references here. However, a full literature list is 
available as a web-based supplement for those who wish to delve more deeply 
into the work performed by the various multiscale simulation and modeling 
communities.

\subsection{Related work}

Aside from numerous publications, project websites and domain-specific reviews,
we have identified a few sources which provide information on multiscale
simulations in various scientific domains. One such source of information is
the {\em Journal of Multiscale Modeling and Simulation} (epubs.siam.org/mms),
which defines itself as an interdisciplinary journal focusing on the
fundamental modeling and computational principles underlying various multiscale
methods. The {\em Journal of Multiscale Modeling} (www.worldscinet.com/jmm/) is
also targeted at multiscale modeling in general.  There are also several books
which present multiscale research in a range of
domains~\cite{Fish:2009,Attinger:2004}, as well as dozens of multiscale
modeling workshops such as the Multiscale Materials Meeting
(www.mrs.org.sg/mmm2012) or the Modelling and Computing Multiscale Systems
workshop (www.computationalscience.nl/MCMS2013). 

There are several articles which focus on the theoretical aspects of multiscale
modelling across domains. Yang et al.~\cite{Yang:2013} present a thorough and
systematic review of the computational and (especially) the conceptual
toolkits for multiscale modelling. In addition, Hoekstra et
al.~\cite{Hoekstra:2010} investigate the modeling aspects of multiscale
simulations, emphasizing simulations using Cellular Automata.

\section{Overview of multiscale communities}

\subsection{Astrophysics}

The astrophysics community hosts a large number of active multiscale projects,
mainly due to the large scale and multi-physics nature of many astrophysical
problems. Because of the intrinsic properties of gravitation,
phenomena on relatively small length scales, e.g. close encounters between
massive stars or galaxies, may have a considerable effect on systems of much
larger size. It is therefore essential in many cases to model these phenomena
using a multiscale approach. Researchers developed multiscale models in a range
of topics of astrophysical interest, such as cosmology\ocite{Springel:2005},
star cluster dynamics\ocite{2008NewA...13..285P,Harfst:2008}, thermonuclear
supernovae\ocite{Ropke:2008} and space weather systems\ocite{Toth:2005}. 
The Space Weather Modeling Framework (http://csem.engin.umich.edu/tools/swmf/index.php) 
is one of the domain-specific toolkits that emerged in this community.

Cactus (www.cactuscode.org\ocite{Goodale:2002}) is a toolkit for coupling simulation codes, which was 
originally used to model black holes, neutron stars and boson stars. Cactus is
now used by researchers in a variety of disciplines, some of which have adopted the
tool to combine single-scale models and construct multiscale simulations.
The Astrophysical Multipurpose Software Environment (AMUSE, www.amusecode.org)
is an extensive and highly versatile toolkit for constructing multiscale
simulations using a wide range of astrophysical codes\ocite{PortegiesZwart:2012}. 
\commentedout{The goal of AMUSE\cite{PortegiesZwart:2012}, as a successor of the MUSE
toolkit~\cite{muse}, is to connect existing (parallel) codes through a layer of
Python and MPI to form coupled simulations, and to allow rapid construction of
multiscale simulations through the use of an intuitive scripting convention.}
AMUSE has been applied, for example, for coupling a gravitational N-body
simulation with a stellar evolution code to model both the dynamical movements
and the aging of stars in a star cluster~\cite{muse}. 
The FLASH 4 code~\cite{Fryxell:2000} combines hydrodynamic solvers with
magnetic field models to simulate the surfaces of compact stars such as white
dwarves and neutron stars. \commentedout{It is highly modular and allows users
to mix and match modules to construct customized multiscale applications.}
Both AMUSE and FLASH\ocite{FLASH} provide extra flexibility by allowing alternative
implementations of its components to co-exist and be interchanged with each
other. They additionally provide simple and elegant mechanisms to
customize code functionalities without requiring modifications to the core
implementation of each component.

\subsection{Biology}

Biological systems, too, span many orders of magnitude through the length and
time scales. Although it is uncommon for researchers to model systems much
larger than the human body (epidemiology is a notable exception), the human body
itself already encompasses many scales, ranging from the molecular scale up to
whole body processes. The sequence from the genome, proteome, metabolome,
physiome to health comprises multi-scale systems biology of the most ambitious
kind\ocite{Noble:2002,Finkelstein:2004,Hetherington:2007,Southern:2008,PittFrancis:2009}.
Multiscale modelling in biology has already been widely reviewed. For example,
Schnell et al.~\cite{Schnell:2007} provide an excellent introduction to the
field, while Dada et al.~\cite{Dada:2010} and Sloot et al.~\cite{Sloot:2010}
respectively provide a general overview of the multiscale modeling efforts in
biology and computational biomedicine. Several coupling tools were originally
developed to construct biomedical multiscale simulations, such as GridSpace
(dice.cyfronet.pl/gridspace) and MUSCLE 2
(http://www.qoscosgrid.org/trac/muscle). In addition, a sizeable number of
markup languages have emerged (e.g., CellML\ocite{Lloyd:2004} and
SBML\ocite{Hucka:2003}) which allow users to exchange definitions of
singlescale models and the system information, an important aspect of
constructing multiscale models.

The Virtual Physiological Human (VPH) Initiative is a large and active
community within the biomedical computing domain. Multiscale simulations and models
have a central role within the VPH, as it supports multiscale modelling 
efforts in Europe (e.g., VPH-NoE, www.vph-noe.eu), USA (e.g., the Multi-scale
Modeling Consortium, www.imagwiki.nibib.nih.gov) as well as world-wide through 
the Physiome project\ocite{Hunter:2008} (www.physiome.org). One recently published 
example involves the coupling of atomistic and continuum subcodes to
model blood flow in the brain~\cite{Grinberg:2012-2}.

\subsection{Energy}

A sizeable number of problems within the energy domain can be resolved using
single-scale models, but especially for nuclear energy problems the use of
multiscale simulations is considered to be fundamentally
important~\cite{Hill:2008}. Modelling a complete nuclear reactor is a highly 
complicated multiscale problem. Here, the testing of both the efficiency and the
durability of reactor parts includes a diverse range of physical
processes that all need to be resolved accurately in computational submodels.
Indeed, a major flaw in one submodel could render the whole reactor ineffective. 
Several tools emerged that assist in coupling fusion applications, such as the
Universal Access Layer (UAL\ocite{Manduchi:2008},
http://www.efda-itm.eu/ITM/html/isip\_ual.html), the Framework Application for
Core-Edge Transport Simulations (FACETS, www.facetsproject.org) and the
Integrated Plasma Simulator (IPS, cswim.org/ips/). Additionally, the developments in
the GriPhyN high energy physics computing project (www.griphyn.org) resulted in
a generalized toolkit for workflow-style multiscale simulations
(Swift~\cite{Wilde:2009}).

As a specific example, the EFDA Task Force on Integrated Tokamak Modeling
(www.efda-itm.eu) is an European initiative which aims to develop a generic yet
comprehensive Tokamak simulator. This simulator can then be applied to
investigate a range of existing and future fusion devices.  The layout of this
simulator is modular and multiscale, including submodels that for example
resolve equilibrium effects, magneto-hydrodynamical stability and heating, with
ab-initio quantum models to be incorporated in the future.

\subsection{Engineering}

Multiscale simulations have been applied to a wide range of engineering
problems, as microscopic properties can be of crucial importance for the
quality of the overall design. In this work, engineering is presented disjoint
from materials science: the former focuses on simulating certain
structures, devices or chemical processes, whereas the latter focuses 
more strongly on the properties of individual materials. 

Fish et al.~\cite{Fish:2009} provide a comprehensive review of the most
commonly used multiscale techniques in the field. Additionally, the {\em
International Journal of Multiscale Computational Engineering}
(http://www.begellhouse.com/journals/multiscale-computational-engineering.html)
has a strong focus on multiscale simulation in engineering.  Multiscale
engineering projects are common within the domain of chemical engineering (see
Lucia et al.~\cite{Lucia:2010} for a comprehensive review), but also include
efforts in aerospace engineering (e.g., DESIDER\ocite{DESIDER} and
FLOMANIA\ocite{FLOMANIA}), non-equilibrium physics\ocite{Martin:2005}, chemical
engineering\ocite{Li:2003}, stochastic simulations of kinetic theory
models\ocite{Cueto:2011} and the coupling of atomistic and continuum methods in
hydrology\ocite{Symeonidis:2005,Wijesinghe:2004}. 

One of the tools that emerged from the engineering domain is the Multiphysics
Object-Oriented Simulation Environment (MOOSE) toolkit
(www.inl.gov/research/moose/). MOOSE is a graphical environment that was
originally used for reactor engineering simulations, but has now been reused
for a range of scientific purposes. A second multiscale coupling environment that recently emerged
from this domain is the Coupled Physics Environment (CouPE, sites.google.com/site/coupempf/).
CouPE allows users to couple different submodels which rely on mesh-based solvers. 

\subsection{Environmental science}

Environmental science covers topics such as ecology studies, climate modeling,
geosciences and hydrology, all of which benefit strongly from multiscale
simulation approaches.  The diverse collection of initiatives include, for
example, hydrology simulations\ocite{vanThang:2010}, weather
forecasting\ocite{Plale:2006,Winglee:2009}, climate modeling\ocite{CCSM} and
disaster predictions\ocite{An:2010}. Klein et al.~\cite{Klein:2011} provide a
broad review of multiscale (fluid dynamics) methods in metereology. Researchers
within this domain have also developed several general-purpose toolkits, such
as the Model Coupling Toolkit\ocite{Larson:2005} (MCT, www.mcs.anl.gov/mct), the Pyre
framework\ocite{Tan:2006} (www.cacr.caltech.edu/projects/pyre), OpenPALM
(www.cerfacs.fr/globc/PALM\_WEB), OASIS (verc.enes.org/oasis),
OpenMI~\cite{Gregersen:2007} and the Bespoke Framework Generator
(BFG, http://cnc.cs.man.ac.uk/projects/bfg.php). The DRIHM project (www.drihm.eu)
aims to develop a distributed research infrastructure, rather than a single
toolkit, to facilitate multiscale hydro-metereological simulations. 

The European Network for Earth System Modelling (www.enes.org) is a large
consortium which is developing a European network for the multiscale modelling
of earth systems. In this consortium the ENSEMBLES project
(ensembles-eu.metoffice.com) uses multiscale ensemble simulations to
simulate the Earth system for climate predictions, which include physical,
chemical, biological and human-related feedback processes. 

\subsection{Materials science}

Materials science applications are inherently multiscale, as the macroscopic properties of 
many materials are largely characterized through interactions occuring on the microscopic
level. Linking our understanding of the physical world at very small scales
with the observable behaviour at the macro-scale is a major focus within this
area of science, and the applications are extremely varied. A popular technique
in this field is coarse-graining, where multiple atoms are resolved as a single 
coarse-grained particle with a pre-imposed potential~\cite{Izvekov:2005}. Several
tools have emerged which facilitate coarse-graining, such as VOTCA (www.votca.org)
and MagiC (code.google.com/p/magic/).

The topics covered in these projects range from multiscale modeling of
radiation damage (e.g., RADINTERFACES\ocite{RADINTERFACES}) to modeling of
multilayered surface systems (e.g., M3-2S\ocite{M3-2S}) and multiscale
heterogeneous modeling of solids~\cite{Weinan:2007}.  The book by Attinger and
Koumoutsakos~\cite{Attinger:2004} comprehensively presents a large number of
projects within the materials sciences.  Additionally, the MMM@HPC project
(www.multiscale-modelling.eu) develops a unified
infrastructure for multiscale materials modelling that covers applications from
first principle quantum mechanics to continuum simulations to model properties
beyond the atomistic scale.  An example of distributed multiscale materials
modeling is the clay-polymer nanocomposites application presented by Suter et
al.~\cite{Suter:2012}\ocite{Groen:2011} Coupling toolkits are relatively
uncommon within this domain, although FEniCS (www.fenicsproject.org) is a tool
that enables multiscale finite-element simulations. 

\subsection{Other communities}

One community of considerable size is the fluid dynamics community, comprising
numerous active areas of research on multiscale simulation. These research
topics include multiscale methods to model multiphase fluids, fluids with
particles\ocite{Garcia:2010,Weinan:2007,Delgado:2003},
biofluids\ocite{Groen:2012,Borgdorff:2012,Bernaschi:2009}, as well as
magnetorheological fluids\ocite{Peng:2011}. The MAPPER project
(www.mapper-project.eu) features several multiscale fluid dynamics
applications, for example to model blood flow and sediment formation in rivers.
The {\em International Journal of Multiscale Computational
Engineering}\ocite{Journal:IJMCE} and the {\em Journal of Multiscale
Modelling}\ocite{Journal:JMM} contain numerous articles on multiscale fluid
dynamics as well.

The multiscale modeling and simulation efforts within fluid dynamics frequently
take place within the context of other scientific domains, such as biology in
the case of blood flow simulations, and environmental science in the case of
river or oceanic simulations. To accommodate this, we have not sought to treat
fluid dynamics as a separate domain, but categorized the projects in
accordance with their application domain.

Overall, the six domains described in this work represent major areas where
multiscale simulations are frequently applied. Having performed an extensive
search, we did find a number of multiscale projects outside these domains. The
vast majority of these projects concern theoretical mathematical modeling of
multiscale problems, and only indirectly relate to the other scientific fields
in our survey. 

\section{Review of multiscale communities}

\begin{figure}[!t]
\centering
\includegraphics[width=2.9in]{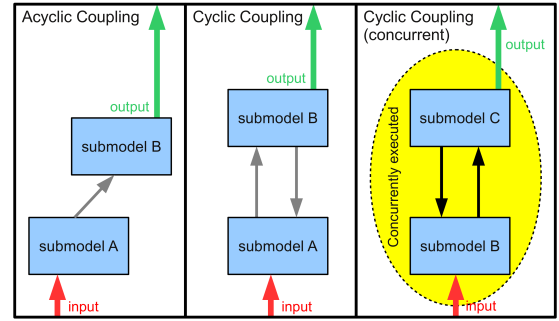}
\caption{Examples of a acyclically (left) and two cyclically coupled (middle
and right) multiscale models. Submodels are indicated by blue boxes, and data
transfers by arrows. On the right we provide a cyclically coupled model where
the submodels are executed concurrently. The concurrent execution is frequently
managed by a software tool that supports cyclic coupling, which we indicate 
there with a yellow ellipse.}
\label{Fig:coupling}
\end{figure}

In this section we characterize several scientific communities, assessing the
prevalence and nature of the multiscale research performed in these domains. We
also review a sizeable number of commonly used multiscale coupling tools, and
reflect on the approaches used in different domains for coupling single-scale
submodels. In our review we distinguish between two multiscale simulation methods:
{\em acyclically coupled} simulations and {\em cyclically coupled} simulations.
Acyclically coupled simulations are applications where subcodes are run,
producing results which in turn are used as input for the execution of a
subsequent subcode. The most characteristic aspect of acyclically coupled
simulations is that there are no cases where two or more subcodes are mutually
dependent of each other during execution. Cyclically-coupled simulations do
have this mutual dependency, and require at least some of the subcodes to be
either run concurrently or in alternating fashion. We show several schematic
examples of multiscale models, both using acyclic coupling and cyclic coupling,
in Fig.~\ref{Fig:coupling}. Although these examples feature two submodels, it is 
not uncommon for multiscale models to consist of three or more different submodels. 

\begin{table*}[!t]
\centering
  \begin{tabular}{c|ccccccc}

  Scientific Domain & Astrophysics
                    & Biology
                    & Energy
                    & Engineering 
                    & Environmental 
                    & Materials\\

  \hline
  Acyclic coupling?                 & some & some & some & most    & many & most \\
  Cyclic coupling?                  & most & most & most & some    & many & some \\
  Concurrent cyclic coupling?       & most & many & many & few     & many & few\\
  Distributed multiscale?           & few  & some & few  & unknown & few  & few \\
  Dominant style of coupling        & D    & G    & D\&G & D       & G    & S\&D \\

\end{tabular}

\caption{Assessed characteristics of the six multiscale simulation domains,
based on the literature we have found. In the last row we list the main style
of submodel coupling used in these disciplines. Here we indicate
domain-specific coupling solutions with a "D", general-purpose
domain-independent solutions with a "G", and collections of hand-written
scripts with an "S". Due to the commercial nature of many engineering
multiscale projects, we are unsure about the dominant style of coupling or the
presence of distributed multiscale simulations in that domain.}

\label{Tab:classification}
\end{table*}

\subsection{Classification of multiscale communities}

We present a brief characterisation of multiscale computing in six scientific
domains in Table~\ref{Tab:classification}. 
Concurrent cyclic coupling is especially common in
astrophysics, and the tight integration of codes required to make concurrent
cyclic coupling possible may be a reason why researchers in this domain tend to
favor custom-tailored domain-specific coupling solutions. Acyclic coupling is
commonly found in the engineering and materials domains, where statistical
averages of smaller-scale simulations are frequently applied to inform
larger-scale models. 

Geographically distributed multiscale simulations are less common, although we
did find at least one example for five of the six domains, and several of them
in biology.  Multiscale efforts in biology, energy and environmental sciences
have resulted in a considerable number of general-purpose coupling tools. We
are unsure why this is the case, but these three domains do all feature large
and internationally coordinated initiatives such as the VPH, ITER and ENES;
organisations which may have been encouraging researchers to adopt generalized
approaches. 

\begin{figure*}[!t]
\centering
\includegraphics[width=5in]{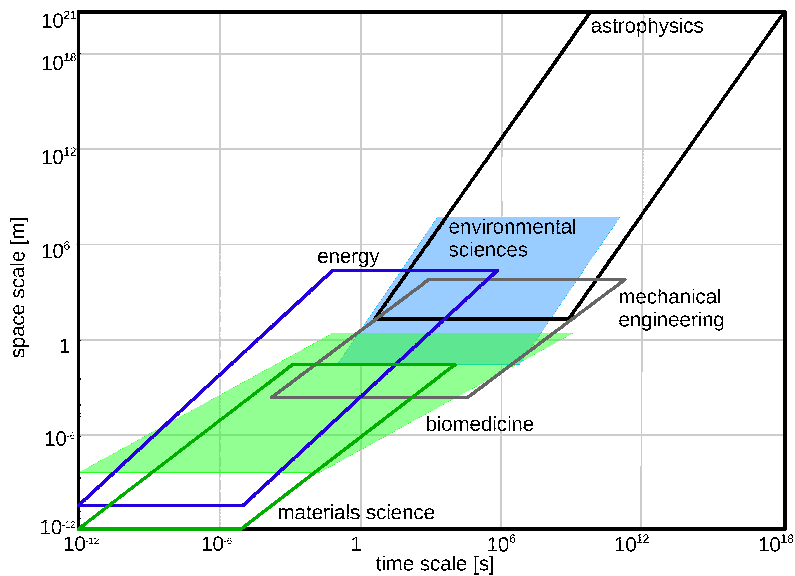}
\caption{Overview of the spatial and temporal scales in which typical (multiscale)
         simulations in several scientific domains operate. Each domain is
         represented as either a colored or a hatched parallelogram.}
\label{Fig:multiscale-all}
\end{figure*}

We present a schematic view of the space and time scales commonly chosen in
different research disciplines in Fig.~\ref{Fig:multiscale-all}.  Each
discipline has a unique scale range given by a parallelogram. For example,
the left-bottom corner of the parallelogram for materials sciences is indicative
of roughly the time steps used in quantum-mechanical studies, while the top-right 
corner is indicative of the duration of mesoscale materials simulations (e.g. 
using finite element methods). Likewise, cosmological dark matter simulations
typically adopt scales which reside at the top end of the astrophysics 
parallelogram. The space and time scale range of each discipline is therefore 
given by the visually observed height and width of the corresponding 
parallelograms. Here, relatively small parallelograms (as seen for mechanical engineering and
environmental science) point to a higher probability of overlapping space 
and/or time scales between subcodes in those disciplines. When scales
between subcodes overlap, cyclic interactions between submodels are essential
to obtain an accurate result, and it becomes difficult to accurately model the
system using acyclic coupling alone. Hoekstra et al.~\cite{Hoekstra:2010}
provide more details on the challenges that arise when scales overlap.
On the other hand, large parallelograms point to a larger range of submodels, 
and an increased likelyhood that three or more submodels are required to solve 
complex problems within these disciplines.

In general, we observe a roughly linear trend between the time scale and the
space scale of simulations across disciplines. This correlation is to be
expected as shorter-range interactions tend to operate on shorter time scales
as well. Additionally, phenomena within a space range between $10^{-4}$m and
$10^4$m and a time range between $10^0$s and $10^4$s are commonly addressed in
many scientific disciplines.  This region of overlap may be particularly
interesting when opting for interdisciplinary approaches or reusable multiscale
simulation tools.  Additionally, when a very high accuracy is required in a
simulation operating on these overlapping scales, it may become increasingly
relevant to incorporate phenomena from other overlapping scientific
disciplines, given that these phenomena are sufficiently proximate. 

\begin{figure}[!t]
\centering
\includegraphics[width=2.2in,angle=270]{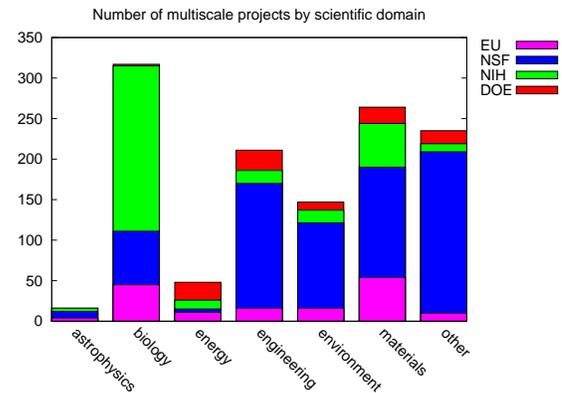}
\caption{Overview of multiscale projects by scientific domain. We obtained the data
from the EU CORDIS database (cordis.europa.eu), the National Institute of Health (projectreporter.nih.gov),
the OSTI database of the Department of Energy (www.osti.gov) and the US National Science Foundation (www.nsf.gov)}
\label{Fig:multiscale-domain}
\end{figure}

\subsection{Prevalence of multiscale research}

To gain some understanding of the size of existing multiscale research communities we have
explored several project data bases from large funding agencies. These
include the European Community Research and Development Information Service
(CORDIS), as well as the project databases of the National Institute for Health (NIH),
the Department of Energy (DOE) and the US National Science Foundation (NSF).
We found the projects by first selecting on the presence of the words 
`multiscale' and `multi-scale' in the project database. For DOE and NIH,
we only selected projects that have these phrases directly in the title, while
we also searched the abstracts in the case of CORDIS and NSF.

\begin{figure}[!t]
\centering
\includegraphics[width=2.2in,angle=270]{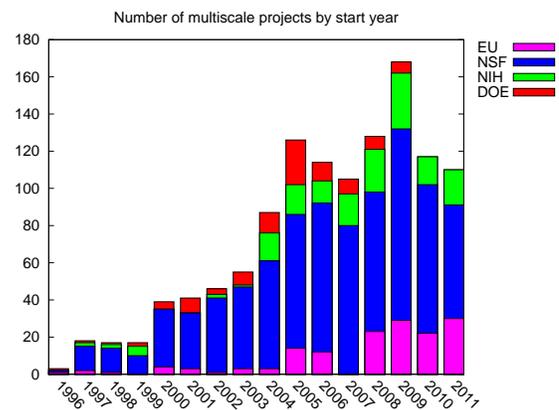}
\caption{Overview of multiscale projects by starting year. We did 
not find any EU Framework project (multiscale or otherwise) which started in 
either 1999 or 2007. Additionally, we found no projects in the DOE database which
were starting in 2010 or 2011.}
\label{Fig:multiscale-year}
\end{figure}

Once we selected the projects, we removed any projects with identical titles,
as these are often continuations of the same project in the previous year.
Also, we eliminated any project that did not describe explicit multiscale
modeling or simulation in its abstract. We found over a thousand multiscale
simulation and modeling grants, which range from multi-million euro
international projects to awards for individual post-doctoral researchers. We
provide an overview of these projects by scientific domain in
Fig.~\ref{Fig:multiscale-domain} and by starting year in
Fig.~\ref{Fig:multiscale-year}. The statistics presented here are by no means
exhaustive, as we only searched for explicit mentions of multiscale and did not
investigate nationally funded projects in the EU, US-based projects funded by
other organizations or projects outside both the EU and the US.  Our results
should therefore be interpreted only as a rough indication of the multiscale
community as a whole and as a lower bound on its size.

In Fig.~\ref{Fig:multiscale-domain} we find that most multiscale projects
reside within the domain of biology and materials, although there are a
considerable number of engineering projects funded in the US. The number of 
EU projects in the astrophysics domain is quite low, most likely because 
international collaboration within theoretical astrophysics tends to focus on
more informal international collaborations and national sources of funding.

In Fig.~\ref{Fig:multiscale-year} we find that multiscale projects emerged in
the late 1990s, and that the number of these projects in the EU has gradually
increased in recent years. The number of multiscale US-based projects peaks in
2009, but has diminished in the last few years. This is in part because the DOE
database contains no projects starting after 2009 (multiscale or otherwise) and
in part because the US Federal Government made a one-time major investment in
scientific research in 2009. As most projects often last three years or more,
we estimate that there are more than 300 multiscale projects currently active.

\begin{figure}[!t]
\centering
\includegraphics[width=2.2in,angle=270]{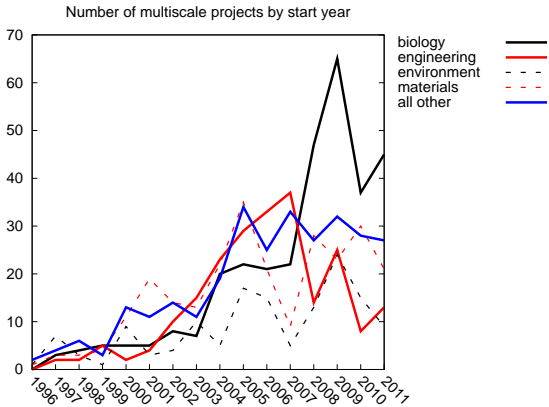}
\caption{Overview of multiscale projects by starting year, separated by domain.
Due to the limited number of projects in energy and astrophysics, we merged these 
domains into the 'other' category.}
\label{Fig:multiscale-domyear}
\end{figure}

We present the number of new projects per year by domain in
Fig.~\ref{Fig:multiscale-domyear}. Here the number of new
multiscale projects in biology is particularly high in 2008 and 2009. This
is largely caused by a growth in funded projects by the EU in 2008
(in part due to the approval VPH projects) and a peak in new
multiscale biology projects funded by NSF and NIH in 2009. The number of
multiscale projects in most other areas has stabilized after 2005, although there
are signs of a decreasing trend in the number of multiscale engineering
projects after 2007. However, as ongoing projects may last as long as 5 years,
we do not know whether the decrease we observe is indeed part of a longer-term
trend.

\subsection{Coupling toolkits for multiscale simulation}

\begin{figure}[!t]
\centering
\includegraphics[width=2.9in]{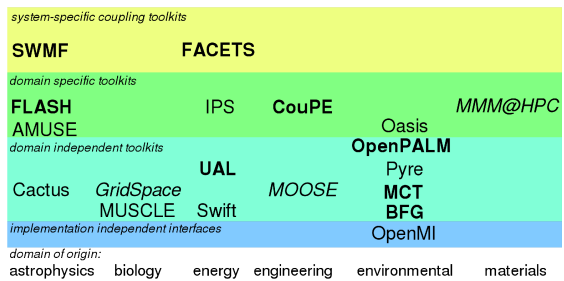}
\caption{Graphical overview of the coupling toolkits discussed in this paper. 
The names of the toolkits are horizontally positioned by their originating domain, and
vertically positioned by their level of generality. Frameworks given in bold font feature a 
user interface based on a compiled language, those in regular font on a scripted
language, and those in cursive font on a graphical user interface.}
\label{Fig:toolkits-generality}
\end{figure}

We classify a large number of coupling toolkits for multiscale simulation in
Table~\ref{Tab:classifySoftware}. Here we indicate whether the tools feature a
generic implementation, intended to be re-used in other domains, what types of
coupling are supported, and whether the tools allow for multiscale simulations
run distributed across multiple computational sites. Allowing the distributed
execution of multiscale simulations is beneficial, because the subcodes within
each simulation may have heterogeneous resource requirements (e.g., some
subcodes may need larger compute resources than others, or require nodes
equipped with specialized hardware). We also provide a graphical overview of
the toolkits along with the originating domain, the type of interface used
and the level of generality in Fig.~\ref{Fig:toolkits-generality}.

In this work we discern several distinct coupling strategies. Perhaps the most
traditional strategy of multiscale coupling is by developing hybrid codes which
cover a set of scales within a single simulation code. These {\em monolithic}
codes are often tailored for specific problems, and can efficiently
incorporate concurrent cyclic coupling for a limited number of built-in submodels.
However, monolithic codes are generally restricted in their modularity and
extensibility. These limitations, combined with the ongoing increase in
available compute capacity, have led to the emergence of more modular and 
flexible coupling approaches, which are easier to extend and refactor but may
have performance limitations when data-intensive concurrent cyclic coupling 
is required. Interestingly, the way different communities have adopted these 
new coupling approaches is not at all uniform.

For example, researchers in astrophysics and energy domains tend to focus on
reusable domain-specific coupling solutions (e.g., AMUSE and IPS), while researchers 
in biology and environmental science focus on general-purpose solutions
(e.g., MUSCLE and OpenPALM). Making a tool general-purpose makes it directly
usable for researchers in other fields, but it may also limit the
functionalities provided by the tool (e.g., lack of unit conversion) or
introduce additional complexity in its architecture to retain flexibility.  We
also provide a brief description of the interface used by the tools, as the
type of interface often provides a useful hint of its intended audience.  Tools
geared towards performance tend to rely often on Fortran and C/C++, tools
geared towards flexibility on Python or Java and tools geared towards
ease-of-use on Graphical User Interfaces (GUIs).  Researchers in the materials
sciences only rarely adopt coupling toolkits, and tend to either employ
inherent multiscale capabilities within molecular dynamics codes (e.g., by
using a ``replica exchange'' method to model a range of temperatures) or to
connect simulations using (often handwritten) pre- and post-processing scripts.
In a few instances, however, they do rely on data conversion libraries such as
the VOTCA toolkit.

Using a single heavyweight and domain-specific toolkit for multiscale
simulations is often convenient for the user in the short term, but it comes
with several drawbacks on the longer term. First, although it is often
straightforward to switch between different solvers within these all-in-one
coupling toolkits (sometimes it is as easy as replacing a single line of
code), it is often much more difficult to switch from one coupling toolkit to
another. This may be necessary if an existing toolkit becomes outdated, or
if the subcodes within that toolkit need to be reused outside of the source 
domain. By constructing and adopting formalizations for defining multiscale
coupling patterns (such as MML~\cite{Borgdorff:2011}), we are able to diminish
this drawback and improve the portability of multiscale simulations and, for
example, allowing them to be more easily moved to a different toolkit if the 
existing one becomes obsolete. 

Another drawback of using traditional all-in-one approaches is that any new
computational improvements in multiscale coupling (such as more powerful data
abstractions or improvements in the data exchange performance between subcodes)
may have to be applied separately to each toolkit to be used to full effect,
resulting in duplicated integration, or even implementation, efforts. This is a
major concern in any large software project, which among other things can be
mitigated by strictly enforcing modularity in the toolkit design (assuming that
the developers of underlying components use standardized APIs that remain
consistent over time).

\begin{table*}[!t]
\centering
  \begin{tabular}{c|ccccccc}
    name
  & domain of origin
  & generic implementation?
  & distributed across sites?
  & acyclic coupling?
  & cyclic coupling?
  & interface presented to users
  & year of last public release\\

  \hline
  AMUSE\ocite{AMUSE}          & astrophysics & no  & yes & yes & yes & Python    & 2013\\
  BFG                         & environment  & yes & no  & yes & yes & Fortran   & 2012\\
  Cactus                      & astrophysics & yes & yes & yes & yes & Custom    & 2012\\
  CouPE                       & engineering  & no  & no  & no  & yes & C++       & 2013\\
  FACETS                      & energy       & no  & n/a & n/a & yes & C++       & 2013\\
  FLASH\ocite{FLASH}          & astrophysics & n/a & n/a & yes & yes & Fortran   & 2013\\
  GridSpace\ocite{GridSpace}  & biology      & yes & yes & yes & n/a & GUI       & 2013\\
  IPS                         & energy       & no  & no  & yes & yes & Python    & not public\\
  MCT\ocite{Larson:2005}      & environment  & yes & yes & yes & yes & Fortran   & 2012\\
  MOOSE Framework             & engineering  & yes & no  & yes & yes & GUI       & not public\\
  MUSCLE\ocite{Hegewald:2008} & biology      & yes & yes & n/a & yes & Java      & 2013\\
  OASIS\ocite{Redler:2010}    & environment  & no  & no  & n/a & yes & Fortran/C & 2012\\
  OpenMI~\cite{Gregersen:2007}& environment  & no  & yes & yes & yes & Java/C\#  & 2011\\
  OpenPALM\ocite{OpenPALM}    & environment  & yes & no  & n/a & yes & GUI       & 2012\\
  Pyre\ocite{Pyre}            & environment  & yes & no  & yes & yes & Python    & 2005\\
  Swift\ocite{Wilde:2011}     & energy       & yes & yes & yes & no  & C-like    & 2012\\ 
  SWMF\ocite{Toth:2005}       & astrophysics & n/a & no  & yes & yes & Fortran   & not public\\
  UAL\ocite{Manduchi:2008}    & energy       & yes & yes & yes & yes & C/Fortran/JAVA & not public\\
\end{tabular}

\caption{Assessed characteristics of the coupling toolkits. All the coupling
toolkits here support the switching and dynamic use of multiple submodels in
a modular way, and the execution of parallel multiscale simulations within a
single compute resource. Within the table we provide a 'yes' if the toolkit
provides this functionality, 'no' if it currently does not appear to do so, and
'n/a' if the functionality appears to be outside of the scope of the toolkit
altogether.} 
\label{Tab:classifySoftware} \end{table*}

\section{Discussion and Conclusions}

We have reviewed a number of multiscale communities and compared them across a
range of criteria. The number of multiscale projects has been increasing in
recent years so that today there are numerous large multiscale projects in a
range of scientific domains. The increase in the number of multiscale projects
also implies a growth in the potential benefit that can be gained by developing
common and reusable multiscale methods. 

The different multiscale communities tend to adopt radically different
technical approaches and possess diverse organizational characteristics.
Within biology, energy and environmental sciences, a considerable fraction of
the multiscale projects are bundled in large international initiatives, while
the multiscale projects within astrophysics and materials sciences are often
driven by much smaller collaborations. On the technical level, researchers in
the astrophysics and energy domains clearly prefer to use domain-specific
toolkits to couple their subcodes, while researchers in biology and
environmental sciences have a stronger inclination towards general-purpose
coupling tools. The numerous projects in the materials sciences adopt yet a
different approach, and frequently construct multiscale simulations by
connecting codes with hand-written scripts. The vast majority of multiscale
simulations are run on single sites, though a small number of projects recently
performed {\em distributed multiscale simulations}, where
individual subcodes are deployed and run on different computational sites.
Considering the heterogeneity in computational requirements of various
subcodes, distributed multiscale simulation may be the only way to efficiently
run production simulations in a number of cases.

In our analysis of scales simulated by different multiscale computing
communities we find a distinct overlap in the scales upon which the simulations
in these domains operate. In particular many research domains feature
simulations on a length scale of about a meter and a time scale of a few hours.
As a result, general-purpose multiscale methods which are geared towards this 
scale may be particularly suitable for reuse by a wide range of scientific 
disciplines, and phenomena operating on these scales in one domain may be of
non-negligible relevance to others.

A uniform strategy for multiscale simulations has yet to emerge, as different
domains have adopted relatively disjoint approaches so far. Nevertheless,
multiscale simulations have become widespread to the point where there are at
least a few hundred active projects in the EU and the US alone.  It is beyond
the scope of this review to fully pronounce on the benefits of pursuing domain
specific approaches versus general purpose approaches for accelerating the
progress of multiscale communities. However, based on the findings we presented
here, we can clearly conclude that it is high time for such an
inter-disciplinary debate to be opened.

\section*{Acknowledgments}

We thank our colleagues in the MAPPER consortium and in particular Krzysztof
Kurowski, Ilya Saverchenko, Kasia Rycerz, Alfons Hoekstra, Marian Bubak, Werner
Dubitzky, Bastien Chopard, David Coster, James Suter and Rupert Nash for their
contributions. We also thank Daniel S. Katz, Oliver Henrich and the
participants of both the Distributed Multiscale Computing workshop 2011 at IEEE
e-Science and the Lorenz Center workshop on Multiscale Modelling and Computing
2013 for their valuable input. This work has received funding from the MAPPER
EC-FP7 project (grant no. RI-261507).



\bibliographystyle{IEEEtran}
\bibliography{Library}

\begin{thebibliography}{10}
\providecommand{\url}[1]{#1}
\csname url@samestyle\endcsname
\providecommand{\newblock}{\relax}
\providecommand{\bibinfo}[2]{#2}
\providecommand{\BIBentrySTDinterwordspacing}{\spaceskip=0pt\relax}
\providecommand{\BIBentryALTinterwordstretchfactor}{4}
\providecommand{\BIBentryALTinterwordspacing}{\spaceskip=\fontdimen2\font plus
\BIBentryALTinterwordstretchfactor\fontdimen3\font minus
  \fontdimen4\font\relax}
\providecommand{\BIBforeignlanguage}[2]{{%
\expandafter\ifx\csname l@#1\endcsname\relax
\typeout{** WARNING: IEEEtran.bst: No hyphenation pattern has been}%
\typeout{** loaded for the language `#1'. Using the pattern for}%
\typeout{** the default language instead.}%
\else
\language=\csname l@#1\endcsname
\fi
#2}}
\providecommand{\BIBdecl}{\relax}
\BIBdecl

\bibitem{Fish:2009}
J.~Fish, \emph{Multiscale Methods: Bridging the Scales in Science and
  Engineering}.\hskip 1em plus 0.5em minus 0.4em\relax Oxford, United Kingdom:
  Oxford University Press, 2009.

\bibitem{Attinger:2004}
S.~Attinger and P.~Koumoutsakos, \emph{Multiscale Modeling and
  Simulation}.\hskip 1em plus 0.5em minus 0.4em\relax Berlin, Heidelberg,
  Germany: Springer-Verlag, 2004.

\bibitem{Yang:2013}
A.~Yang, ``On the common conceptual and computational frameworks for multiscale
  modeling,'' \emph{Industrial \& Engineering Chemistry Research (in press)},
  2013.

\bibitem{Hoekstra:2010}
A.~Hoekstra, A.~Caiazzo, E.~Lorenz, J.-L. Falcone, and B.~Chopard,
  \emph{Complex Automata: Multi-scale Modeling with Coupled Cellular
  Automata}.\hskip 1em plus 0.5em minus 0.4em\relax Springer-Verlag, 2010, pp.
  29--57.

\bibitem{Springel:2005}
V.~{Springel}, ``{The cosmological simulation code GADGET-2},'' \emph{\mnras},
  vol. 364, pp. 1105--1134, Dec. 2005.

\bibitem{2008NewA...13..285P}
S.~{Portegies Zwart}, S.~{McMillan}, D.~{Groen}, A.~{Gualandris}, M.~{Sipior},
  and W.~{Vermin}, ``{A parallel gravitational N-body kernel},'' \emph{New
  Astronomy}, vol.~13, pp. 285--295, Jul. 2008.

\bibitem{Harfst:2008}
S.~{Harfst}, A.~{Gualandris}, D.~{Merritt}, and S.~{Mikkola}, ``{A hybrid
  N-body code incorporating algorithmic regularization and post-Newtonian
  forces},'' \emph{\mnras}, vol. 389, pp. 2--12, Sep. 2008.

\bibitem{Ropke:2008}
F.~K. {R{\"o}pke} and R.~{Bruckschen}, ``{Thermonuclear supernovae: a
  multi-scale astrophysical problem challenging numerical simulations and
  visualization},'' \emph{New Journal of Physics}, vol.~10, no.~12, pp.
  125\,009--+, Dec. 2008.

\bibitem{Toth:2005}
G.~{T{\'o}th}, I.~V. {Sokolov}, T.~I. {Gombosi}, D.~R. {Chesney}, C.~R.
  {Clauer}, D.~L. {De Zeeuw}, K.~C. {Hansen}, K.~J. {Kane}, W.~B. {Manchester},
  R.~C. {Oehmke}, K.~G. {Powell}, A.~J. {Ridley}, I.~I. {Roussev}, Q.~F.
  {Stout}, O.~{Volberg}, R.~A. {Wolf}, S.~{Sazykin}, A.~{Chan}, B.~{Yu}, and
  J.~{K{\'o}ta}, ``{Space Weather Modeling Framework: A new tool for the space
  science community},'' \emph{Journal of Geophysical Research (Space Physics)},
  vol. 110, no.~A9, p. A12226, dec 2005.

\bibitem{Goodale:2002}
\BIBentryALTinterwordspacing
T.~Goodale, G.~Allen, G.~Lanfermann, J.~Mass{\'o}, T.~Radke, E.~Seidel, and
  J.~Shalf, ``The {Cactus} framework and toolkit: Design and applications,'' in
  \emph{Vector and Parallel Processing -- VECPAR'2002, 5th International
  Conference, Lecture Notes in Computer Science}.\hskip 1em plus 0.5em minus
  0.4em\relax Berlin: Springer, 2003. [Online]. Available:
  \url{http://edoc.mpg.de/3341}
\BIBentrySTDinterwordspacing

\bibitem{PortegiesZwart:2012}
S.~{Portegies Zwart}, S.~{McMillan}, I.~{Pelupessy}, and A.~{van Elteren},
  ``{Multi-physics Simulations using a Hierarchical Interchangeable Software
  Interface},'' in \emph{Advances in Computational Astrophysics: Methods,
  Tools, and Outcome}, ser. Astronomical Society of the Pacific Conference
  Series, R.~{Capuzzo-Dolcetta}, M.~{Limongi}, and A.~{Tornamb{\`e}}, Eds.,
  vol. 453, Jul. 2012, p. 317.

\bibitem{muse}
S.~{Portegies Zwart}, S.~{McMillan}, S.~{Harfst}, D.~{Groen}, M.~{Fujii},
  B.~{\'{O} Nuall\'{a}in}, E.~{Glebbeek}, D.~{Heggie}, J.~{Lombardi}, P.~{Hut},
  V.~{Angelou}, S.~{ Banerjee}, H.~{Belkus}, T.~{Fragos}, J.~{Fregeau},
  E.~{Gaburov}, R.~{Izzard}, M.~{Juric}, S.~{Justham}, A.~{Sottoriva},
  P.~{Teuben}, J.~{van Bever}, O.~{Yaron}, and M.~{Zemp}, ``A multiphysics and
  multiscale software environment for modeling astrophysical systems,''
  \emph{New Astronomy}, vol.~14, no.~4, pp. 369 -- 378, 2009.

\bibitem{Fryxell:2000}
B.~Fryxell, K.~Olson, P.~Ricker, F.~X. Timmes, M.~Zingale, D.~Q. Lamb,
  P.~MacNeice, R.~Rosner, J.~W. Truran, and H.~Tufo, ``Flash: An adaptive mesh
  hydrodynamics code for modeling astrophysical thermonuclear flashes,''
  \emph{The Astrophysical Journal Supplement Series}, vol. 131, no.~1, p. 273,
  2000.

\bibitem{FLASH}
\BIBentryALTinterwordspacing
``Flash,'' 2011. [Online]. Available: \url{FLASH multiscale framework:
  http://flash.uchicago.edu/website/home/}
\BIBentrySTDinterwordspacing

\bibitem{Noble:2002}
D.~Noble, ``Modeling the heart--from genes to cells to the whole organ,''
  \emph{Science}, vol. 295, no. 5560, pp. 1678--1682, 2002.

\bibitem{Finkelstein:2004}
A.~Finkelstein, J.~Hetherington, L.~Li, O.~Margoninski, P.~Saffrey, R.~Seymour,
  and A.~Warner, ``Computational challenges of systems biology,''
  \emph{Computer}, vol.~37, pp. 26--33, May 2004.

\bibitem{Hetherington:2007}
J.~Hetherington, I.~D.~L. Bogle, P.~Saffrey, O.~Margoninski, L.~Li,
  M.~Varela~Rey, S.~Yamaji, S.~Baigent, J.~Ashmore, K.~Page, R.~M. Seymour,
  A.~Finkelstein, and A.~Warner, ``Addressing the challenges of multiscale
  model management in systems biology,'' \emph{Computers and Chemical
  Engineering}, vol.~31, no.~8, pp. 962 -- 979, 2007.

\bibitem{Southern:2008}
J.~Southern, J.~Pitt-Francis, J.~Whiteley, D.~Stokeley, H.~Kobashi, R.~Nobes,
  Y.~Kadooka, and D.~Gavaghan, ``Multi-scale computational modelling in biology
  and physiology,'' \emph{Progress in Biophysics and Molecular Biology},
  vol.~96, no. 1-3, pp. 60 -- 89, 2008.

\bibitem{PittFrancis:2009}
J.~Pitt-Francis, P.~Pathmanathan, M.~O. Bernabeu, R.~Bordas, J.~Cooper, A.~G.
  Fletcher, G.~R. Mirams, P.~Murray, J.~M. Osborne, and A.~e.~a. Walter,
  ``Chaste: A test-driven approach to software development for biological
  modelling,'' \emph{Computer Physics Communications}, vol. 180, no.~12, pp.
  2452 -- 2471, 2009.

\bibitem{Schnell:2007}
S.~Schnell, R.~Grima, and P.~K. Maini, ``Multiscale modeling in biology,''
  \emph{American Scientist}, vol.~95, pp. 134--142, 2007.

\bibitem{Dada:2010}
\BIBentryALTinterwordspacing
J.~O. Dada and P.~Mendes, ``Multi-scale modelling and simulation in systems
  biology,'' \emph{Integr. Biol.}, vol.~3, pp. 86--96, 2011. [Online].
  Available: \url{http://dx.doi.org/10.1039/C0IB00075B}
\BIBentrySTDinterwordspacing

\bibitem{Sloot:2010}
P.~M.~A. Sloot and A.~G. Hoekstra, ``Multi-scale modelling in computational
  biomedicine,'' \emph{Briefings in Bioinformatics}, vol.~11, no.~1, pp.
  142--152, 2010.

\bibitem{Lloyd:2004}
C.~M. Lloyd, M.~D.~B. Halstead, and P.~F. Nielsen, ``Cellml: its future,
  present and past,'' \emph{Progress in Biophysics and Molecular Biology},
  vol.~85, no. 2–3, pp. 433 -- 450, 2004.

\bibitem{Hucka:2003}
M.~Hucka, A.~Finney, H.~M. Sauro, H.~Bolouri, J.~C. Doyle, and H.~e.~a. Kitano,
  ``The systems biology markup language (sbml): a medium for representation and
  exchange of biochemical network models,'' \emph{Bioinformatics}, vol.~19,
  no.~4, pp. 524--531, 2003.

\bibitem{Hunter:2008}
P.~J. Hunter, E.~J. Crampin, and P.~M.~F. Nielsen, ``Bioinformatics, multiscale
  modeling and the iups physiome project,'' \emph{Briefings in Bioinformatics},
  vol.~9, no.~4, pp. 333--343, 2008.

\bibitem{Grinberg:2012-2}
L.~Grinberg, J.~Insley, D.~Fedosov, V.~Morozov, M.~Papka, and G.~Karniadakis,
  ``Tightly coupled atomistic-continuum simulations of brain blood flow on
  petaflop supercomputers,'' \emph{Computing in Science Engineering}, vol.~14,
  no.~6, pp. 58--67, 2012.

\bibitem{Hill:2008}
D.~J. {Hill}, ``{Nuclear energy for the future},'' \emph{Nature Materials},
  vol.~7, pp. 680--682, Sep. 2008.

\bibitem{Manduchi:2008}
G.~Manduchi, F.~Iannone, F.~Imbeaux, G.~Huysmans, J.~Lister, B.~Guillerminet,
  P.~Strand, L.-G. Eriksson, and M.~Romanelli, ``A universal access layer for
  the integrated tokamak modelling task force,'' \emph{Fusion Engineering and
  Design}, vol.~83, no. 2-3, pp. 462 -- 466, 2008.

\bibitem{Wilde:2009}
\BIBentryALTinterwordspacing
M.~Wilde, I.~Foster, K.~Iskra, P.~Beckman, Z.~Zhang, A.~Espinosa, M.~Hategan,
  B.~Clifford, and I.~Raicu, ``Parallel scripting for applications at the
  petascale and beyond,'' \emph{Computer}, vol.~42, pp. 50--60, November 2009.
  [Online]. Available: \url{http://dx.doi.org/10.1109/MC.2009.365}
\BIBentrySTDinterwordspacing

\bibitem{Lucia:2010}
A.~Lucia, ``Multi-scale methods and complex processes: A survey and look
  ahead,'' \emph{Computers \& Chemical Engineering}, vol.~34, no.~9, pp. 1467
  -- 1475, 2010.

\bibitem{DESIDER}
\BIBentryALTinterwordspacing
``{DESIDER: detached eddy simulations for industrial aerodynamics.}'' 2011.
  [Online]. Available: \url{http://cfd.mace.manchester.ac.uk/desider/}
\BIBentrySTDinterwordspacing

\bibitem{FLOMANIA}
\BIBentryALTinterwordspacing
``{FLOMANIA: flow physics modelling, an integrated approach.}'' 2011. [Online].
  Available: \url{http://cfd.mace.manchester.ac.uk/flomania/}
\BIBentrySTDinterwordspacing

\bibitem{Martin:2005}
D.~F. Martin, P.~Colella, M.~Anghel, and F.~J. Alexander, ``Adaptive mesh
  refinement for multiscale nonequilibrium physics,'' \emph{Computing in
  Science and Engineering}, vol.~7, pp. 24--31, 2005.

\bibitem{Li:2003}
J.~Li and M.~Kwauk, ``Exploring complex systems in chemical engineering—the
  multi-scale methodology,'' \emph{Chemical Engineering Science}, vol.~58, no.
  3–6, pp. 521 -- 535, 2003.

\bibitem{Cueto:2011}
E.~Cueto, M.~Laso, and F.~Chinesta, ``Meshless stochastic simulation of
  micro-macrokinetic theory models,'' \emph{International Journal of Multiscale
  Computational Engineering}, vol.~9, no.~1, pp. 1--16, 2011.

\bibitem{Symeonidis:2005}
V.~Symeonidis, G.~E. Karniadakis, and B.~Caswell, ``A seamless approach to
  multiscale complex fluid simulation,'' \emph{Computing in Science and
  Engineering}, vol.~7, pp. 39--46, 2005.

\bibitem{Wijesinghe:2004}
H.~S. Wijesinghe and N.~G. Hadjiconstantinou, ``Discussion of hybrid
  atomistic-continuum methods for multiscale hydrodynamics,''
  \emph{International Journal for Multiscale Computational Engineering},
  vol.~2, no.~2, 2004.

\bibitem{vanThang:2010}
P.~{van Thang}, B.~{Chopard}, L.~{Lef{\`e}vre}, D.~A. {Ondo}, and E.~{Mendes},
  ``{Study of the 1D lattice {B}oltzmann shallow water equation and its
  coupling to build a canal network},'' \emph{Journal of Computational
  Physics}, vol. 229, pp. 7373--7400, Sep. 2010.

\bibitem{Plale:2006}
B.~Plale, D.~Gannon, J.~Brotzge, K.~Droegemeier, J.~Kurose, D.~McLaughlin,
  R.~Wilhelmson, S.~Graves, M.~Ramamurthy, R.~D. Clark, S.~Yalda, D.~A. Reed,
  E.~Joseph, and V.~Chandrasekar, ``Casa and lead: Adaptive cyberinfrastructure
  for real-time multiscale weather forecasting,'' \emph{IEEE Computer},
  vol.~39, pp. 56--64, 2006.

\bibitem{Winglee:2009}
R.~M. {Winglee}, E.~{Harnett}, and A.~{Kidder}, ``{Relative timing of substorm
  processes as derived from multifluid/multiscale simulations: Internally
  driven substorms},'' \emph{Journal of Geophysical Research (Space Physics)},
  vol. 114, no. A13, p. A09213, sep 2009.

\bibitem{CCSM}
\BIBentryALTinterwordspacing
``Community climate system model,'' 2011. [Online]. Available:
  \url{http://www.scidac.gov/CCSM/}
\BIBentrySTDinterwordspacing

\bibitem{An:2010}
C.~An and Y.~Cai, ``The effect of beach slope on the tsunami run-up induced by
  thrust fault earthquakes,'' \emph{Procedia Computer Science}, vol.~1, no.~1,
  pp. 645 -- 654, 2010, iCCS 2010.

\bibitem{Klein:2011}
R.~Klein, S.~Vater, E.~Paeschke, and D.~Ruprecht, \emph{Multiple scales methods
  in meteorology}.\hskip 1em plus 0.5em minus 0.4em\relax Springer Verlag,
  2011, pp. 127--196.

\bibitem{Larson:2005}
J.~Larson, R.~Jacob, and E.~Ong, ``The model coupling toolkit: A new
  {Fortran90} toolkit for building multiphysics parallel coupled models,''
  \emph{International Journal of High Performance Computing Applications},
  vol.~19, no.~3, pp. 277--292, Fall 2005.

\bibitem{Tan:2006}
E.~Tan, E.~Choi, P.~Thoutireddy, M.~Gurnis, and M.~Aivazis, ``Geoframework:
  Coupling multiple models of mantle convection within a computational
  framework,'' \emph{Geochem. Geophys. Geosyst.}, vol.~7, no. Q06001, 2006.

\bibitem{Gregersen:2007}
J.~B. Gregersen, P.~J.~A. Gijsbers, and S.~J.~P. Westen, ``{OpenMI: Open
  modelling interface},'' \emph{Journal of Hydroinformatics}, vol.~9, no.~3,
  pp. 175--191, 2007.

\bibitem{Izvekov:2005}
S.~Izvekov and G.~A. Voth, ``A multiscale coarse-graining method for
  biomolecular systems,'' \emph{The Journal of Physical Chemistry B}, vol. 109,
  no.~7, pp. 2469--2473, 2005.

\bibitem{RADINTERFACES}
\BIBentryALTinterwordspacing
``{RADINTERFACES},'' 2011. [Online]. Available:
  \url{http://www.materials.imdea.org/Research/Projects/RADINTERFACE/tabid/342%
3/Default.aspx}
\BIBentrySTDinterwordspacing

\bibitem{M3-2S}
\BIBentryALTinterwordspacing
``M3-2s,'' 2011. [Online]. Available: \url{http://www.m3-2s.bham.ac.uk/}
\BIBentrySTDinterwordspacing

\bibitem{Weinan:2007}
E.~Weinan, B.~Engquist, X.~Li, W.~Ren, and E.~Vanden-Eijnden, ``{Heterogeneous
  Multiscale Methods: A Review},'' \emph{Communications in Computational
  Physics}, vol.~2, no.~3, pp. 367--450, Jun. 2007.

\bibitem{Suter:2012}
J.~Suter, D.~Groen, L.~Kabalan, and P.~V. Coveney, ``Distributed multiscale
  simulations of clay-polymer nanocomposites,'' in \emph{Materials Research
  Society Spring Meeting}, vol. 1470.\hskip 1em plus 0.5em minus 0.4em\relax
  San Francisco, CA: MRS Online Proceedings Library, April 2012.

\bibitem{Groen:2011}
D.~Groen, J.~Suter, and P.~V. Coveney, ``Modelling distributed multiscale
  simulation performance: An application to nanocomposites,'' in \emph{Seventh
  IEEE international conference on e-Science and Grid computing: Stockholm,
  Sweden}.\hskip 1em plus 0.5em minus 0.4em\relax Piscataway, NJ: IEEE Computer
  Society, December 2011, pp. 105--111.

\bibitem{Garcia:2010}
A.~Donev, J.~B. Bell, A.~L. Garcia, and B.~J. Alder, ``A hybrid
  particle-continuum method for hydrodynamics of complex fluids,''
  \emph{Multiscale Model. Simul.}, vol.~8, pp. 871--911, 2010.

\bibitem{Delgado:2003}
R.~Delgado-Buscalioni and P.~V. Coveney, ``Continuum-particle hybrid coupling
  for mass, momentum, and energy transfers in unsteady fluid flow,''
  \emph{Phys. Rev. E}, vol.~67, p. 046704, Apr 2003.

\bibitem{Groen:2012}
D.~Groen, J.~Hetherington, H.~B. Carver, R.~W. Nash, M.~O. Bernabeu, and P.~V.
  Coveney, ``Analyzing and modeling the performance of the {HemeLB}
  lattice-{B}oltzmann simulation environment,'' \emph{submitted to JoCS}, 2012.

\bibitem{Borgdorff:2012}
J.~Borgdorff, C.~Bona-Casas, M.~Mamonski, K.~Kurowski, T.~Piontek, B.~Bosak,
  K.~Rycerz, E.~Ciepiela, T.~Guba{\l}a, and D.~e.~a. Harezlak, ``{A Distributed
  Multiscale Computation of a Tightly Coupled Model Using the Multiscale
  Modeling Language},'' \emph{Procedia Computer Science}, vol.~9, pp. 596--605,
  2012.

\bibitem{Bernaschi:2009}
M.~{Bernaschi}, S.~{Melchionna}, S.~{Succi}, M.~{Fyta}, E.~{Kaxiras}, and J.~K.
  {Sircar}, ``{MUPHY: A parallel MUlti PHYsics/scale code for high performance
  bio-fluidic simulations},'' \emph{Computer Physics Communications}, vol. 180,
  pp. 1495--1502, Sep. 2009.

\bibitem{Peng:2011}
Y.-B. Peng and J.~Li, ``Multiscale analysis of stochastic fluctuations of
  dynamic yield of magnetorheological fluids,'' \emph{International Journal for
  Multiscale Computational Engineering}, vol.~9, no.~2, pp. 175--191, 2011.

\bibitem{Journal:IJMCE}
\BIBentryALTinterwordspacing
``International journal for multiscale computational engineering,'' 2012.
  [Online]. Available:
  \url{http://www.begellhouse.com/journals/61fd1b191cf7e96f.html}
\BIBentrySTDinterwordspacing

\bibitem{Journal:JMM}
\BIBentryALTinterwordspacing
``Journal of multiscale modelling,'' 2012. [Online]. Available:
  \url{http://www.worldscinet.com/jmm/}
\BIBentrySTDinterwordspacing

\bibitem{Borgdorff:2011}
J.~Borgdorff, J.-L. Falcone, E.~Lorenz, B.~Chopard, and A.~G. Hoekstra, ``{A
  principled approach to distributed multiscale computing, from formalization
  to execution},'' in \emph{Proceedings of the IEEE 7th International
  Conference on e-Science Workshops}.\hskip 1em plus 0.5em minus 0.4em\relax
  Stockholm, Sweden: IEEE Computer Society Press, 2011, pp. 97--104.

\bibitem{AMUSE}
\BIBentryALTinterwordspacing
``Amuse, the astrophysical multipurpose software environment,'' 2011. [Online].
  Available: \url{http://www.amusecode.org}
\BIBentrySTDinterwordspacing

\bibitem{GridSpace}
\BIBentryALTinterwordspacing
``Distibuted computing environments - gridspace technology,'' 2011. [Online].
  Available: \url{http://dice.cyfronet.pl/gridspace/}
\BIBentrySTDinterwordspacing

\bibitem{Hegewald:2008}
J.~Hegewald, M.~Krafczyk, J.~T\"{o}lke, A.~Hoekstra, and B.~Chopard, ``An
  agent-based coupling platform for complex automata,'' in \emph{Proceedings of
  the 8th international conference on Computational Science, Part II}, ser.
  ICCS '08.\hskip 1em plus 0.5em minus 0.4em\relax Berlin, Heidelberg:
  Springer-Verlag, 2008, pp. 227--233.

\bibitem{Redler:2010}
R.~{Redler}, S.~{Valcke}, and H.~{Ritzdorf}, ``{OASIS4 - a coupling software
  for next generation earth system modelling},'' \emph{Geoscientific Model
  Development}, vol.~3, pp. 87--104, Jan. 2010.

\bibitem{OpenPALM}
\BIBentryALTinterwordspacing
``{OpenPALM},'' 2012. [Online]. Available:
  \url{http://www.cerfacs.fr/globc/PALM\_WEB/}
\BIBentrySTDinterwordspacing

\bibitem{Pyre}
\BIBentryALTinterwordspacing
``The pyre framework,'' 2012. [Online]. Available:
  \url{http://www.cacr.caltech.edu/projects/pyre/}
\BIBentrySTDinterwordspacing

\bibitem{Wilde:2011}
M.~Wilde, M.~Hategan, J.~M. Wozniak, B.~Clifford, D.~S. Katz, and I.~Foster,
  ``Swift: {A} language for distributed parallel scripting,'' \emph{Parallel
  Computing}, vol.~39, no.~9, pp. 633--652, September 2011.

\end{thebibliography}

\section*{Biographies}

\subsection*{Derek Groen}
Derek Groen is a Post-doctoral Research Associate in the CCS at
University College London, specialised in multiscale simulation and
parallel/distributed computing. He has worked with a wide range of
applications, including those using lattice-Boltzmann, N-body and molecular
dynamics methods, and participated in several EU-funded IT projects. He
finished his PhD in 2010 at the University of Amsterdam, where he investigated
the performance of N-body simulations run across geographically distributed
(supercomputing) infrastructures. Derek currently works on constructing and
testing multiscale simulation applications of cerebrovascular bloodﬂow and
clay-polymer nanocomposites.

\subsection*{Stefan J. Zasada}
Stefan J. Zasada is a software engineer in the Centre for Computational Science
at UCL. He develops lightweight grid middleware and enabling tools for
``e-Science''. Stefan has a Masters degree in Advanced Software Engineering from
the University of Manchester, where he was responsible for implementing the
WS-Security specification in Perl for use by the WSRF::Lite toolkit. He is
currently the lead developer on the AHE project and involved in developing a
toolkit for the Virtual Physiological Human Initiative. In addition, he is
working on a PhD in Computer Science, investigating the design and development
of lightweight grid middleware and resource allocation solutions. 

\subsection*{Peter V. Coveney}
Prof. Peter V. Coveney holds a Chair in Physical Chemistry and is Director
of the Centre for Computational Science, and an Honorary Professor in
Computer Science. He is also Professor Adjunct within the Medical School at
Yale University, and Director of the UCL Computational Life and Medical
Sciences Network. Coveney is active in a broad area of interdisciplinary
theoretical research including condensed matter physics and chemistry,
materials science, life and medical sciences. He has published over 300 papers,
books and edited works, including the acclaimed bestsellers The Arrow of
Time and Frontiers of Complexity, both with Roger Highfield.

\end{document}